\documentstyle[preprint,revtex]{aps}
\draft

\begin{document}
\begin{title}
Disordered Boson Systems: A Perturbative Study
\end{title}
\author{Lizeng Zhang}
\begin{instit}
Department of Physics and Astronomy, \\
The University of Tennessee, Knoxville,
Tennessee 37996-1200
\end{instit}
\begin{abstract}
A hard-core disordered boson system is mapped onto a quantum spin
$1/2$ XY-model with transverse random fields. It is then
generalized to a system of spins with an arbitrary magnitude $S$ and
studied through a $1/S$ expansion. The first order $1/S$ expansion
corresponds to a spin-wave theory. The effect of weak disorder is
studied perturbatively within such a first order $1/S$ scheme.
We compute the reduction of the speed of sound and the
life time of the Bloch phonons in the regime of weak disorder.
Generalizations of the present study to the strong
disordered regime are discussed.
\end{abstract}

\narrowtext

\section{Introduction}

Recent experiments on liquid $He^{4}$ in vycor and other
porous media \cite{EXPT} have revealed some interesting
features of disordered boson systems. For a boson system
at low temperature, strong disorder may destroy
superfluidity and transforms it onto a Bose-glass phase.
While many theoretical efforts have been devoted to
this disorder-tuned quantum critical phenomenon
\cite{MHL}$-$\cite{FGG}, some important issues are
yet to be explored and to be understood. Like it in
the fermion system, study of the nature of low energy
excitations is extremely important to
the understanding of the macroscopic properties.
In this paper we study the properties of low energy
excitations in weakly disordered boson systems and
their effects on superfluidity. This is of interest
both to compare to the pure case, and to study the
precursor to the destruction of superfluidity.
Although the experiment on superconductor-insulator
transition in dirty films \cite{Liu} is an extremely
interesting possible realization of disordered boson
systems \cite{FGG}, the present paper will be
restricted to the study of dirty bosons with
short-ranged interaction, which is suited to model
systems like $He^{4}$ in random media.

As usual in theoretical physics, one may wish to proceed
the study by starting with a perturbation theory about
some known limiting cases. However, it is clear that the
non-interacting system is not a suitable starting point,
since disorder will cause a (artificial) Bose-Einstein
condensation of bosons onto the lowest energy, hence
{\em localized} state. This condensate is unstable with
respect to arbitrary weak but finite repulsions
between bosons. Therefore it is essential to
include the interaction between bosons to
proceed the study of the effect of disorder. Since the
condensate is no longer uniform in the presence of disorder,
the self-consistency required for the normal Hartree theory
imposes a set of non-linear equations on the condensate
which are not solvable in general. In this paper, we adopt
an alternative approach. We start by considering a model
of lattice hard-core bosons with random potential,
introduced by Ma, Halperin and Lee \cite{MHL}:
\begin{equation}\label{hhc}
{\cal H} = -t\sum_{<i,j>}b^{+}_{i}b_{j} + H.c +
\sum_{j}(W_{j}-\mu) b^{+}_{j}b_{j} \;\;,
\end{equation}
where $b^{+}_{j}$, $b_{j}$ are boson creation and annihilation
operators at lattice site $j$, with the hard-core constraint
$b^{+}_{j}b_{j} = 0,1$.
$<i,j>$ indicates nearest neighbor and $W_{j}$ is the random on-site
potential obeys certain (independent) distribution.
This model is equivalent to a quantum spin $1/2$ XY-model with
transverse random fields \cite{MHL}:
\begin{equation}\label{hrso}
{\cal H} = -J\sum_{<i,j>}(S_{i}^{x}S_{j}^{x} +
S_{i}^{y}S_{j}^{y}) - \sum_{j} h_{j} S_{j}^{z} \;\; ,
\end{equation}
with
$$
S^{+}_{j} \; \longleftrightarrow \; b^{+}_{j}\;\;, \;\;\;\;\;
S^{-}_{j} \; \longleftrightarrow \; b^{-}_{j}\;\;,
$$
$$
J \; \longleftrightarrow \; 2t \;\;, \;\;\;\;\;
h_{j} \; \longleftrightarrow \; \mu - W_{j} \;\;.
$$
Superfluidity in the boson model (\ref{hhc}) corresponds to a
spin long range order in the XY plane. Thinking of the boson
problem in terms of the spin language and vice versa is proven
to be a fruitful way in understanding the physics of these
problems \cite{MHL,ZM1,ZM2}. For the most part of the paper,
we shall devote ourselves to investigate the ground state
properties.

Above mapping is an exact mathematical transformation. Now we
apply approximations in term of the spin language. We start by
letting the magnitude of the spins to have an arbitrary (integer or
half integer) value $S$, roughly corresponding to relaxing the hard
core condition and letting the site occupation number of bosons
$b^{+}_{j} b_{j} = 0,1, \ldots , 2S$.
When $S \rightarrow \infty$, (\ref{hrso})
describes a system of classical spins and the usual
mean field theory becomes exact. We study this
Hamiltonian  through a $1/S$ expansion. When finite
$S$ is considered, one introduces quantum fluctuations
as well as stronger on-site repulsions between bosons.
To the first order of $1/S$, this expansion describes
the Gaussian fluctuations and corresponds to a generalized
spin wave theory, where the magnons correspond to the
phonon excitations in the boson system. Since in the
classical case the long range order persists up to the
infinitely strong disorder limit (provided that the
probability distribution for the random field
$h_{j}$, $P(h_{j})$, is finite at its mean value)
\cite{MHL}, the expansion is assumed to be inside the
superfluid phase. In the end, we will discuss the
possibility of probing the transition to the disordered
phase within this approach.

Our present study is confined within the first order $1/S$
expansion, i.e., the spin wave theory. In this scheme, the system is
described by a quadratic Hamiltonian of (the Holstein-Primakoff)
bosons with zero chemical potential. For the pure system, which
will be discussed in detail in the following section, this theory
gives the same results as those in the Bogoluibov theory for weakly
interacting bosons. We study the disordered system in section III.
In this case, the random transverse fields $\{h_{j}\}$ result in
both diagonal and off-diagonal disorder for the Holstein-Primakoff
bosons. For weak randomness, where perturbation theory is applicable,
modifications to the low energy spectrum due to disorder are obtained
analytically. We find that in this regime the superfluidity is
rather robust against (weak) disorder, as one expects on the
physical ground. We conclude our discussions in section IV.

\section{Pure System}

In this section we present the spin wave theory results for the pure
system. Consider the quantum XY-model (\ref{hrso}) without fields
\begin{equation}\label{hp0}
{\cal H}^{0} = -J\sum_{<i,j>}(S_{i}^{z}S_{j}^{z} +
S_{i}^{x}S_{j}^{x}) \;\;,
\end{equation}
where we have made a spin rotation for convenience. Assuming the
ground  state to be ferromagnetic with broken symmetry in the
$z$-direction, the classical ground state is then given
by $<S^{z}_{i}> = S$. Fluctuations
about the classical ground state can be studied through the usual
Holstein-Primakoff transformation \cite{HP}, given by
\begin{eqnarray}\label{hpt}
S_{j}^{+} = \sqrt{2S}(1-a_{j}^{\dagger}a_{j} / 2S)^{1/2}a_{j} \;\; ,
\nonumber \\
S_{j}^{-} = \sqrt{2S} a_{j}^{\dagger}
(1-a_{j}^{\dagger}a_{j} / 2S)^{1/2} \;\; , \nonumber \\
S_{j}^{z} = S-a_{j}^{\dagger}a_{j} \;\; ,
\end{eqnarray}
where $a_{j}^{\dagger}$, $a_{j}$ are bosonic creation and
annihilation operators. Assuming periodic boundary condition
on a $d$-dimensional hypercubic lattice, in terms of the
Fourier transformed variables
\begin{equation}
a_{k}^{\dagger} = \frac{1}{\sqrt{N}}\sum_{j}
e^{-ikx_{j}}a_{j}^{\dagger} \;\;,\;\;\;\;
a_{k} = \frac{1}{\sqrt{N}}\sum_{j} e^{ikx_{j}}a_{j} \;\;,
\end{equation}
we have, with rescaling $J \rightarrow J/S^{2}$,
\begin{equation}\label{hpak}
{\cal H}^{0} = -\frac{z}{2}NJ +
\frac{zJ}{S}\sum_{k}a_{k}^{\dagger}a_{k} -
\frac{zJ}{4S} \sum_{k}\gamma_{k}(a_{k}^{\dagger}a_{k} +
a_{k} a_{-k} + H.c. ) + {\cal O}(\frac{1}{S^{3/2}}) \;\; ,
\end{equation}
where
\begin{equation}
\gamma_{k} = \frac{1}{z} \sum_{\delta} e^{ik\delta} \;\; .
\end{equation}
$\delta$ above is the directional unit vector of the lattice, and $z$
is the lattice coordination number. Ignoring the terms of higher power
in $1/S$, the usual Bogoluibov transformation gives
\begin{equation}
{\cal H}^{0} = E^{0} + \sum_{k} {\omega}^{0}_{k} b_{k}^{+} b_{k}
\;\; ,
\end{equation}
with
\begin{equation}
E^{0} = \frac{zJ}{2S}\sum_{k}(\sqrt{1-\gamma_{k}}-1)-
\frac{z}{2}NJ \;\; ,
\end{equation}
\begin{equation}\label{omega0}
\omega^{0}_{k} = \frac{zJ}{S}\sqrt{1-\gamma_{k}} \;\; .
\end{equation}
Here $b^{+}_{k}$, $b_{k}$ are the new creation and annihilation
operators respectively, with
\begin{equation}
b_{k} = u_{k}a_{k} - v_{k}a_{-k}^{+} \;\;, \;\;\;\;
b_{k}^{+} = -v_{k}a_{-k} + u_{k}a_{k}^{+} \;\;,
\end{equation}
and
\begin{equation}\label{transf}
u_{k}^{2} = \frac{1}{2} \left (
\frac{1-\frac{\gamma_{k}}{2}}{\sqrt{1-\gamma_{k}}} + 1
\right ) \;\;, \;\;\;\;
v_{k}^{2} = \frac{1}{2} \left (
\frac{1-\frac{\gamma_{k}}{2}}{\sqrt{1-\gamma_{k}}} - 1
\right ) \;\;, \;\;\;\;
2u_{k}v_{k} = \frac{\frac{\gamma_{k}}{2}}{\sqrt{1-\gamma_{k}}} \;\;.
\end{equation}
The low energy excitations are magnetic
phonons (magnons) with speed of sound $c_{s}^{0} = \sqrt{z}J/S$.

To have some physical feelings about the nature of the low energy
excitations, we express the magnon operators in terms of the spin
operators (to the leading order of $1/S$):
\begin{equation}
b_{k}=\frac{1}{\sqrt{2S}} \left ( i(1-\gamma_{k})^{-1/4}S_{k}^{y} -
(1-\gamma_{k})^{1/4}S_{k}^{x} \right )
\;\; ,
\end{equation}
\begin{equation}
b^{+}_{k}=\frac{1}{\sqrt{2S}} \left (
-i(1-\gamma_{k})^{-1/4}S_{k}^{y} - (1-\gamma_{k})^{1/4}S_{k}^{x}
\right ) \;\; ,
\end{equation}
where $\vec{S}_{k}$ is the Fourier transform of $\vec{S}_{j}$.
$S_{k}^{x}$ and $S_{k}^{y}$ generate out-plane oscillations and
in-plane rotations respectively. Above decomposition of the magnon
operators shows that while high energy excitations are composed of
both kind spin motions, the low energy, long wavelength excitations
are dominantly in-plane rotations. The zero mode
($\omega_{k=0}^{0} = 0$) is a pure in-plane uniform rotation.
As we will see later, such a zero mode also exists in the
disordered system.

To calculate the helicity modulus, which is proportional
to the superfluid density \cite{FBJ}, we impose a phase twist
on the order parameter. Let us maintain the periodic boundary
condition in the first $d-1$ directions but impose the anti-periodic
boundary condition in the $d$-th direction for
the in-plane component of the spins. Since the spin wave
excitations are well defined only with the correct (classical)
ground state, we need to determine first the classical ground state
under such a boundary condition. It is clear that the spin
configuration which gives the lowest energy under
such circumstance is the state in which spins are rotated gradually
from angle $0$ to $\pi$ in the $d$-th direction (See the next section
for discussions for general situations.). Under such a rotation,
spins in each $d-1$ dimensional plane perpendicular to the axis of
the phase twist are rotated by the same amount, and the angular
difference of the rotations between two successive planes is $\pi/L$,
where $L$ is the linear size of the lattice. To apply the
Holstein-Primakoff
transformation (\ref{hpt}), we perform therefore a spin rotation
around the $y$-axis (in the spin space), and the angle of rotation on
the spin at site $j = (j_{1},....,j_{d-1},j_{d})$ is
\begin{equation}
\phi_{j} = \frac{\pi j_{d}}{L} \;\; .
\end{equation}
In terms of the rotated spins, (\ref{hp0}) with anti-periodic
boundary condition becomes
\begin{equation}\label{hap0}
{\cal H}^{0} = -J\sum_{<i,j>} \left [
cos(\phi_{i}-\phi_{j})(S_{i}^{z}S_{j}^{z} +
S_{i}^{x}S_{j}^{x})
-sin(\phi_{i}-\phi_{j})(S_{i}^{z}S_{j}^{x} - S_{i}^{x}S_{j}^{z})
\right ] \;\; .
\end{equation}
The spin variables in (\ref{hap0}) satisfy the periodic
boundary condition, and they are ordered in the $z$-direction
in the ground state.

Applying the Holstein-Primakoff transformation (\ref{hpt}), and
through a procedure similar to that in the case of the periodic
boundary condition, we have, to the order $1/S$,
\begin{equation}
{{\cal H}'}^{0}={E'}^{0}+\sum_{k}{\omega'}^{0}_{k} b_{k}^{+}b_{k}
\;\; ,
\end{equation}
with
\begin{eqnarray}
{E'}^{0} = \frac{J}{2S}\sum_{k} ( \sqrt{(z-2+2\alpha ) [(z-
2+2\alpha)-(z-2)\gamma_{k}^{\perp} - 2\alpha\gamma^{\parallel}_{k}]}
\nonumber \\
- (z-2+2\alpha) ) -\frac{z-2}{2}NJ - \alpha NJ  \;\; ,
\end{eqnarray}
\begin{equation}
{\omega'}^{0}_{k} = \frac{J}{S}\sqrt{(z-2+2\alpha )
[(z-2+2\alpha)-(z-2)\gamma_{k}^{\perp} -
2\alpha\gamma^{\parallel}_{k}]}  \;\; ,
\end{equation}
$$
\alpha = cos(\frac{\pi}{L}) \;\; ,\;\;\; \gamma_{k}^{\parallel} =
\frac{1}{2} \sum_{\delta_{\parallel}} e^{ik\delta_{\parallel}} \;\;,
\;\;\; \gamma_{k}^{\perp} =
\frac{1}{z-2} \sum_{\delta_{\perp}} e^{ik\delta_{\perp}} \;\;,
$$
and $\parallel$, $\perp$ are defined with respect to the axis of phase
twist.

Compare above results with those obtained with the periodic boundary
condition, we see that, by twisting the phase of the order parameter,
one lifts the ground state energy of the system but lowers the
excitation energy (with respect to the new ground state energy level),
i.e., ${E'}^{0}>E^{0}$, ${\omega'}_{k}^{0} < \omega_{k}^{0}$. Since
to the first order of $1/S$ the second term in (\ref{hap0}) vanishes,
we can think of the effect of changing the boundary condition as a
reduction of the spin coupling in the $d$-th
direction, i.e., $J \rightarrow J$ in the first $d-1$ directions and
$J \rightarrow \alpha J$ in the $d$-th direction. Because of this
reduction, one loses ground state energy but makes excitations
easier. Moreover, since this coupling reduction is anisotropic,
the system picks up an easy direction of excitation
along the $d$-axis.

The helicity modulus $\gamma$ \cite{FBJ} can now be easily
calculated from the expressions obtained above:
\begin{equation}
\gamma (T) = \lim_{L \rightarrow \infty} 2(\frac{L}{\pi})^{2}
\frac{\Delta F}{N}
= \lim_{L \rightarrow \infty} 2(\frac{L}{\pi})^{2}
\frac{1}{N} \left (\Delta E^{0} + \frac{N}{(2\pi)^{d}} \int d^{d}k
\frac{\Delta\omega_{k}^{0}}{e^{\beta\omega_{k}^{0}} -1} \right )
\;\; ,
\end{equation}
where $\Delta F$ refers to the change of free energy as a result of
changing boundary condition and $\Delta E^{0} = {E'}^{0}-E^{0}$,
$\Delta\omega_{k}^{0} = {\omega'}_{k}^{0} - \omega_{k}^{0}
\propto (\frac{\pi}{L})^{2}k$. This gives
\begin{equation}\label{gmT}
\gamma(T) = \gamma(0) - aT^{d+1} \;\; ,
\end{equation}
where $a$ is a positive constant. This is in agreement with the known
result for weakly interacting boson systems \cite{LLP}.

Before we go further, we would like to comment on the validity
of the spin wave approach to our problem. In one dimension (1D),
the quantum correction to
the in-plane magnetization, $\delta m \equiv \frac{1}{N} \sum_{j}
<a_{j}^{\dagger}a_{j}>$, is divergent \cite{SWP}, reflecting the lack of
{\em true} long range order in the quantum XY model \cite{LSM}.
For two and higher dimensions, it has been shown rigorously that
finite magnetization exists in the XY plane \cite{KLS},
and indeed the quantum correction to the classical magnetization
in the spin wave theory becomes finite also. The fact that
$\delta m$ decreases with increasing dimensionality indicates
that spin wave theory becomes a better approximation in
higher dimensions. Notice that despite the divergence
in $\delta m$, and thus the absence of the condensate, the
superfluidity that exists (at zero temperature) in one
dimension manifests itself
through the linear behavior in low energy excitations.

Finally, we give the spin wave solution for the XY model in
a uniform transverse field where $h_{j} \equiv h_{0}$. This
can be obtained in a similar manner to the procedure for
the case of $h_{0} = 0$ shown above, with a rescaling
$J \rightarrow J/S^{2}$, $h_{0} \rightarrow h_{0}/S$.
Comparison of the results
obtained here with those calculated in the presence of the
random fields to be described in the next section will help us
isolate the effect of disorder. For large field, i.e., for
$h_{0}\geq zJ$, the classical spins are
completely aligned with the transverse field and therefore there
is no in-plane magnetization. For weaker field, the classical
ground state (with the periodic boundary condition) is given
by $<S_{j}^{z}> = Scos\theta_{0}$, $<S_{j}^{y}> = Ssin\theta_{0}$,
where the transverse field $h_{0}$ tilts the (classical) spins with
an angle $\theta_{0}$ such that $sin\theta_{0} = H_{0} \equiv
h_{0}/zJ$. The ground state energy
\begin{equation}
E^{0}(h_{0}) = \frac{zJ}{2S}\sum_{k}\sqrt{(1-\gamma_{k})
(1-\gamma_{k} H_{0}^{2})}-\frac{zJ}{2}N(1 + H_{0}^{2}) \;\; .
\end{equation}
The magnon spectrum is modified by $h_{0}$ as
\begin{equation}\label{omegah0}
\omega_{k}^{0}(h_{0}) = \frac{zJ}{S}\sqrt{(1-\gamma_{k})
(1-\gamma_{k} H_{0}^{2})} \;\; .
\end{equation}
One can compute the out-plane susceptibility $\chi_{\perp}$
(compressibility in the boson language) by taking
second derivative of $E^{0}$:
\begin{equation}
\chi_{\perp} = \frac{\partial^{2} E^{0}/N}{\partial
(h_{0}/S)^{2}}|_{h_{0}=0} =\frac{S^{2}}{zJ} + {\cal O}(S) \;\; .
\end{equation}
The $S^{2}$ term is simply the classical result, and the rest are
corrections due to quantum fluctuations. Since the classical
contribution to $\gamma(0)$ is just $J$, we calculate the speed of
sound
\begin{equation}
c_{s}^{2} = \frac{\gamma (0)}{\chi_{\perp}} = \frac{zJ^{2}}{S^{2}}
\;\; ,
\end{equation}
in agreement with the result obtained earlier.
We see that from pure classical considerations one can obtain
$c_{s}$ which characterizes the quantum nature of the low energy
excitations.

Not surprisingly the speed of sound is reduced by $h_{0}$, with
$c_{s}^{0}(h_{0}) = c_{s}^{0}\sqrt{1-H^{2}_{0}}$. However, it is less
obvious that the (relative) quantum correction to the in-plane
magnetization
\begin{equation}
\frac{\delta m(h_{0})}{Scos\theta_{0}} = \frac{1}{NS}\sum_{k}
<a_{k}^{\dagger}a_{k}> = \frac{1}{2S}\int \frac{d^{d}k}{(2\pi)^{d}}
\left ( \frac{1-\frac{\gamma_{k}}{2}(1+H^{2}_{0})}{\sqrt{(1-\gamma_{k})
(1-\gamma_{k} H^{2}_{0})}} - 1 \right ) \;\; ,
\end{equation}
is also reduced, i.e., $\frac{\delta m(h_{0})}{Scos\theta_{0}} <
\frac{\delta m(0)}{S}$. This suggests that the reduction of
$<S_{j}^{z}>$ by a uniform  transverse field is mainly a classical
effect, resulted from the tilting of spins off the ZX (XY in the
original coordinate system) plane by $h_{0}$. In fact, as
$H_{0} \rightarrow 1$, so that $cos\theta_{0} \rightarrow 0$,
$\delta m/Scos\theta_{0} \rightarrow 0$, which helps to explain
why the critical value of $H_{0}$ is just the classical
value. We will compare this with the effect of random fields
as we proceed.

In summary, the quantum XY model provides us a useful representation
of the boson problem in studying the physics of superfluidity.
In the large $S$ theory, the expansion parameter $1/S$ contains
both the effect of the repulsive interactions between
bosons and the effect of quantum zero point fluctuations.
We have shown that for the pure system the quadratic
fluctuations represented in the first order $1/S$ (spin wave)
theory gives the same physics as that described in the Bogoluibov
theory for weakly interacting bosons. Now we apply this method
to study the disordered case.

\section{Disordered System}

Consider the disordered system (with periodic boundary
condition) given by
\begin{equation}\label{hr0}
{\cal H} = -\frac{J}{S^{2}}\sum_{<i,j>}(S_{i}^{z}S_{j}^{z} +
S_{i}^{x}S_{j}^{x}) - \sum_{j}\frac{h_{j}}{S} S_{j}^{y} \;\; ,
\end{equation}
with
\begin{equation}\label{rfdist}
\overline{h_{j}} = 0 \;\;,\;\;\; \overline{h_{i}h_{j}} =
\delta_{ij}h^{2} \;\; .
\end{equation}
In (\ref{hr0}), we have made the rescaling $J \rightarrow J/S^{2}$,
$h_{j} \rightarrow h_{j}/S$. The overline indicates average over
the random transverse fields. The choice of $\overline{h_{j}} = 0$
corresponds to a `particle-hole symmetry' in the boson system. For
hard-core bosons, it corresponds to an average occupancy of half
particle per site. Assume that spins in the ground state of
the pure (classical) system are ordered in the $z$-direction,
the transverse random fields along the $y$-axis then tilt
each (classical) spin according to the equation
of motion (with $\vec{S}_{j}^{classical} = S(0,sin\theta_{j},
cos\theta_{j})$ ):
\begin{equation}\label{ceom}
sin\theta_{j}J\sum_{<j'>}cos\theta_{j'} = h_{j}cos\theta_{j}
\;\; ,
\end{equation}
where $<j'>$ indicates nearest neighbors of the site $j$. Now we
perform a local rotation of the (quantum) spin $\vec{S}_{j}$ over
an angle $\theta_{j}$ about the $x$-axis so that the classical
ground state is uniformly ordered in the $z$-axis in terms of
the rotated spins. After the rotation (\ref{hr0}) becomes
\begin{eqnarray}\label{hrsaa}
{\cal H} = -\frac{J}{S^{2}}\sum_{<i,j>}(S_{i}^{x}S_{j}^{x} +
cos\theta_{i}
cos\theta_{j}S_{i}^{z}S_{j}^{z} + sin\theta_{i}sin\theta_{j}
S_{i}^{y}S_{j}^{y} \nonumber \\
- sin\theta_{i}cos\theta_{j}S_{i}^{y}S_{j}^{z} -
cos\theta_{i}sin\theta_{j} S_{i}^{z}S_{j}^{y}) -
\sum_{j}\frac{h_{j}}{S}(sin\theta_{j}S_{j}^{z} +
cos\theta_{j}S_{j}^{y}) \;\; .
\end{eqnarray}
Applying the Holstein-Primakoff transformation (\ref{hpt}) now,
together with (\ref{ceom}), one has
\begin{eqnarray}\label{hrrs}
{\cal H} = -J\sum_{<i,j>}cos\theta_{i}cos\theta_{j}
-\sum_{j}h_{j}sin\theta_{j} -\frac{J}{2S}\sum_{<i,j>}
[(1-sin\theta_{i}sin\theta_{j})a_{i}a_{j} + \nonumber \\
(1+sin\theta_{i}sin\theta_{j})a_{i}^{\dagger}a_{j} + H.c ] +
\frac{1}{S}\sum_{j}\frac{h_{j}}{sin\theta_{j}}a_{j}^{\dagger}a_{j} +
{\cal O}(\frac{1}{S^{3/2}}) \;\; .
\end{eqnarray}

Rewrite (\ref{hrrs}) in terms of Fourier transformed
variables, we have
\begin{equation}
{\cal H}={\cal H}^{0} + {\cal H}^{1} \;\; ,
\end{equation}
with ${\cal H}^{0}$ given by (\ref{hpak}) and
\begin{equation}
{\cal H}^{1} = E^{10} + \sum_{k,k'}u_{k-k'}a_{k}^{\dagger} a_{k'}
+\sum_{k,k'}(v_{-k,-k'}a_{k} a_{k'} -
v_{k,-k'}a_{k}^{\dagger} a_{k'} + H.c ) \;\; ,
\end{equation}
where
\begin{equation}
E^{10} = J \sum_{<i,j>}(1-cos\theta_{i}cos\theta_{j}) - \sum_{j}
h_{j}sin\theta_{j} \;\; ,
\end{equation}
\begin{equation}
u_{k} = \frac{1}{NS} \sum_{j} (\frac{h_{j}}{sin\theta_{j}} -zJ)
e^{ikx_{j}} \;\; ,
\end{equation}
\begin{equation}
v_{k,k'} = \frac{J}{2NS} \sum_{<i,j>} sin\theta_{i}sin\theta_{j}
e^{i(kx_{i}+k'x_{j})} \;\; .
\end{equation}
In the weak disorder limit, ${\cal H}^{1}$ may be considered as a
perturbation to ${\cal H}^{0}$.

In stead of carrying out a Bogoluibov type transformation,
which becomes non-local and depends on the configuration of
the random fields in the disordered case, we compute poles
of the propagator $\overline{<{\cal T}[a_{k}(t)a_{k}^{\dagger}(0)]>}$
($<\ldots >$ denotes the ground state expectation value and
${\cal T}$ indicates time ordering.). These poles are directly
related to the excitation spectrum of the system. To this
end, we consider
\begin{equation}
F(k,k';t) \equiv -i \left ( \begin{array}{c}
<{\cal T}[a_{k}(t)a_{k'}^{+}(0)]>  \\
<{\cal T}[a_{-k}^{+}(t)a_{k'}^{+}(0)]>
\end{array} \right ) \;\; ,
\end{equation}
which obeys the equation of motion
\begin{equation}
(i\frac{\partial}{\partial t} - T_{k})F(k,k';t) =
\delta(t)\delta_{kk'}E_{1} +\sum_{k"}U_{k,k"}F(k",k';t) \;\; ,
\end{equation}
where
\begin{equation}
T_{k} = \frac{zJ}{S} \left ( \begin{array}{cc}
1 - \frac{\gamma_{k}}{2} & - \frac{\gamma_{k}}{2}  \\
\frac{\gamma_{k}}{2}     & -(1 - \frac{\gamma_{k}}{2})
\end{array} \right ) \;\; ,\;\;\;\;\;
E_{1} = \left ( \begin{array}{c}
1  \\ 0
\end{array} \right ) \;\; ,
\end{equation}
and
\begin{equation}
U_{k,k'} = \left ( \begin{array}{cc}
u_{k-k'}-v_{k,-k'}-v_{-k',k}  & v_{k,-k'}+v_{-k',k}  \\
-(v_{k,-k'}+v_{-k',k})        & -(u_{k-k'}-v_{k,-k'}-v_{-k',k})
\end{array} \right ) \;\; .
\end{equation}
Applying the Fourier transformation in time, with
\begin{equation}
F(k,k';\omega) = \int_{-\infty}^{\infty} dt F(k,k';t)e^{i\omega t}
\;\; ,
\end{equation}
we have
\begin{equation}
(\omega - T_{k})F(k,k';\omega) = \delta_{kk'}E_{1} + \sum_{k"}U_{k,k"}
F(k",k';\omega) \;\; ,
\end{equation}
or
\begin{equation}\label{Fkkp}
F(k,k';\omega) =(\omega - T_{k})^{-1} \delta_{kk'}E_{1} +
(\omega - T_{k})^{-1} \sum_{k"}U_{k,k"} F(k",k';\omega) \;\; .
\end{equation}
For the pure system, $U_{k,k'} = 0$, we have
\begin{equation}
F^{0}(k,k';\omega) = \tilde{F}^{0}(k,\omega)\delta_{k,k'} E_{1}
\equiv G^{0}(k,\omega) \;\; ,
\end{equation}
with
\begin{equation}
\tilde{F}^{0}(k,\omega) \equiv (\omega - T_{k})^{-1} =
\frac{1}{D^{0}(k,\omega)}
\left ( \begin{array}{cc}
\omega + \frac{zJ}{S}(1 - \frac{\gamma_{k}}{2}) &
-\frac{zJ}{S}\frac{\gamma_{k}}{2}
\\
\frac{zJ}{S}\frac{\gamma_{k}}{2}     &
\omega-\frac{zJ}{S}(1 - \frac{\gamma_{k}}{2})
\end{array} \right ) \;\; ,
\end{equation}
and
\begin{equation}\label{det0}
\frac{1}{D^{0}(k,\omega)} =
\frac{1}{(\omega-\omega_{k}^{0}+i\eta)(\omega+\omega_{k}^{0}-i\eta)}
\;\;,\;\;\;\;\; \eta \rightarrow 0^{+} \;\;.
\end{equation}
This gives
\begin{equation}
G^{0}(k,\omega) =
\left ( \begin{array}{c}
\frac{u_{k}^{2}}{\omega - \omega_{k}^{0} + i\eta} -
\frac{v_{k}^{2}}{\omega + \omega_{k}^{0} - i\eta}
\\ u_{k}v_{k} \left ( \frac{1}{\omega-\omega_{k}^{0}+i\eta} -
\frac{1}{\omega+\omega_{k}^{0}-i\eta} \right )
\end{array} \right ) \;\; ,
\end{equation}
where $u_{k}$ and $v_{k}$ are given by (\ref{transf}). Thus,
from the poles in $G^{0}(k,\omega)$, we have recovered the
excitation spectrum obtained earlier through the Bogoluibov
transformation.

For disordered system, one can iterate (\ref{Fkkp}) to find a
perturbative solution for $F(k,k';\omega)$. $U_{k,k'}$ contains
randomness that needs to be averaged out. It depends on the random
fields $\{h_{j}\}$ both explicitly and implicitly in $\{\theta_{j}\}$
through (\ref{ceom}). For weak randomness, where $h<<zJ$, one can
solve (\ref{ceom}) order by order (see the appendix). One has, in the
regime of weak disorder,
\begin{equation}\label{Gf}
G(k,\omega) \equiv \overline{F(k,k;\omega)} =
(\tilde{F}^{0-1}(k,\omega) -
\Sigma^{*}(k,\omega))^{-1} E_{1}
\end{equation}
where the self energy $\Sigma^{*}(k,\omega)$ is given by
\begin{equation}
\Sigma^{*}(k,\omega) = \overline{U_{k,k}} + \sum_{k'}
\overline{U_{k,k'}\tilde{F}^{0}(k',\omega)U_{k',k}} \;\; .
\end{equation}
The detailed algebraic expression of $\Sigma^{*}(k,\omega)$
is given in the appendix.
This is consistent with a perturbation expansion for the self energy
to the order $h^{4}$ (see the appendix). Notice that in real space
$U_{k,k'}$ is not just a random on-site potential. It contains both
(correlated) diagonal and off-diagonal disorder.

The poles of $G(k,\omega)$ are given by (\ref{poleG}) when
the distribution of the random fields is Gaussian, which can
be computed analytically for 1D system. As discussed earlier,
while the divergence in $\delta m$ indicates the instability
of the classical ground state and thus the absence of true long
range order in the quantum system, the linear behavior in
$\omega_{k}^{0}$ at small $k$ indicates the existence of the
superfluidity at zero temperature. It is therefore meaningful
to investigate deviations from $\omega_{k}^{0}$ due to disorder
in 1D systems within the present approach. To the order
$h^{4}$, the poles of $G(k,\omega)$ are given by (with $z=2$)
\begin{equation}\label{spect1d}
\omega = \omega_{k} =
\omega_{k}^{0} [1 - \frac{1}{2} (\frac{h}{zJ})^{4} A(k)] \;\; ,
\end{equation}
with
\begin{equation}\label{real1}
{\em Re}\{A\} = \frac{1}{4} + 5cosk - 3cos^{3}k + cos^{4}k \;\;,
\end{equation}
\begin{equation}
{\em Im}\{A\} = \frac{(2-cosk)^{2}}{2}|sink|(1 + cosk) +
(1 + cos^{2}k) \frac{1 - cosk}{4|sink|} \;\; .
\end{equation}
We see that the linear mode persists at low energy with a
reduced speed of sound
$c_{s}=c_{s}^{0}(1-\frac{13}{8}(\frac{h}{zJ})^{4})$. It has been
shown \cite{ZM1} that for $S=1/2$ any amount of disorder will
change the power law behavior of the spin-spin correlation
function to an exponential one, corresponding to the instability of
superfluidity in the 1D hard-core boson system. For systems with
soft-core bosons, which are roughly described by (\ref{hrso}) with
$S>1/2$, renormalization group
study \cite{GS} shows that superfluidity may persist in the presence
of (weak) disorder, if the value of an exponent $\eta$ of the
correlation function in the corresponding pure system is less than
a critical value $\eta_{c}=1/3$. The fact that in our calculations
the linear mode persists in
the weak disorder limit suggests that present $1/S$ theory
describes systems with $S>S_{c}$ such that the exponent
$\eta < \eta_{c}$. The imaginary part of the pole diverges at
the zone boundary ($k=\pm \pi$), indicating vanishing life time of
these (Bloch) modes. By analyzing the scattering rate of the Bloch
phonons using Fermi's golden rule, this divergence can be interpreted
as the divergence of the density of states in one dimension.
In 2D, the imaginary part has a (van Hove) singularity at the band
center, and it behaves regularly in higher dimensions.

For an arbitrary dimension $d$, the correction to the speed
of sound is given by
\begin{equation}\label{csr}
c_{s}=c^{0}_{s} \left (1-\frac{1}{2} (\frac{h}{zJ})^{4}A(0)\right)
\;\; ,
\end{equation}
with
\begin{equation}
A(0) = \frac{3z+3}{2z} + {\cal P}\int\frac{d^{d}k}{(2\pi)^{d}}
\frac{\frac{1}{d^{2}}\sum_{j=1}^{d}sin^{2}k_{j}}{1-\gamma_{k}} \;\; .
\end{equation}
The imaginary part of
$D(k,\omega) \equiv det(\tilde{F}^{0-1} - \Sigma^{*})$
vanishes for small $k$ and $\omega$ as $\omega^{d+2}$
(see the appendix), implying a decay rate of phonons $\tau^{-1}
\propto \omega^{d+1}$, and the mean free path
$l=c_{s}\tau \propto k^{-(d+1)}$. Thus for small
$k$, $kl >>1$. This implies that momentum is still
approximately conserved. We see that both the
reduction of the sound speed and the decay of phonons
become weaker in higher dimensions. Notice that this is
not a time decay rate for phonons, which can only be
nonzero through multi-phonon scattering, but is just the scattering
rate for a Bloch phonon. Since the scattering is elastic, energy
conservation implies $\tau^{-1} \propto \omega^{d-1}$, the extra
suppression factor of $\omega^{2}$ is presumably a reflection that
there always exists a zero frequency mode which corresponds to uniform
rotation in the ordering plane. Since for small $\omega $,
$\omega\tau >>1$, the Bloch modes remain robust in the presence
of weak disorder. The fact that the zero mode ($\omega_{k=0} = 0$)
has zero imaginary part shows that it remains to be an exact
eigenmode of the Hamiltonian in the disordered system.

Next we study the effect of disorder on the magnitude of
the order parameter, which corresponds to the square root
of the condensate density in the boson language. It is given by
\begin{equation}
m = \overline{<S^{z}_{i}>} = \frac{1}{N}\overline{\sum_{j}
cos\theta_{j}(S - <a_{j}^{\dagger}a_{j}>)} \;\; ,
\end{equation}
which can simply be understood as following:
the first term gives the reduction in $m$ due to the tilt of the
classical spins from the ordering plane. The second term shows that
each Holstein-Primakoff boson lowers the spin along $z'$-axis by one,
and hence only by $cos\theta_{j}$ along the $z$-axis. Thus to
the order $h^{2}$
\begin{eqnarray}
m = \frac{1}{N}\sum_{j}\overline{cos\theta_{j}} (S -
\frac{1}{N}\sum_{k}<a_{k}^{\dagger}a_{k}>_{0})  \nonumber \\
= (1-\frac{1}{2}(\frac{h}{zJ})^{2})
\left ( S-\frac{1}{2}\int\frac{d^{d}k}{(2\pi)^{d}} (\frac{1-
\frac{\gamma_{k}}{2}}{\sqrt{1-\gamma_{k}}}-1) \right ) \;\;,
\end{eqnarray}
where $<a_{k}^{\dagger}a_{k}>_{0}$ is the expectation value
with respect to ${\cal H}^{0}$ (since
$\overline{\delta<a_{k}^{\dagger}a_{k}>}$ vanishes
to order $h^{2}$, see (\ref{dm}).).

Again let $m=m_{classical}-\delta m$, where $m_{classical}$ is the
classical ($S \rightarrow \infty$) condensate, and $\delta m$ is the
reduction due to quantum fluctuations. In the pure case where
$h_{j} = h_{0}$ and $\theta_{j} = \theta_{0}$, we have seen that
besides the `classical' factor $cos\theta_{0}$, $\delta m$ is
also reduced by $H_{0}$, resulting in a net reduction of
$\delta m/m_{classical}$ by a uniform field. This is in contrast
to that in the random system considered here, where $\delta m /
m_{classical}$ is independent of the disorder to order $h^{2}$.
While $\delta m$ is reduced by disorder to order $h^{2}$, indicating
that disorder and quantum fluctuations have opposing effects on the
condensate density, it will not be the case to higher order in $h$,
and that $\delta m$ will also be enhanced by disorder. To order
$h^{4}$, we have (see the appendix)
\begin{eqnarray}\label{qfd}
\frac{\delta m}{m_{classical}} = \frac{1}{2S}\int
\frac{d^{d}k}{(2\pi)^{d}}
\left ( \frac{1-\frac{\gamma_{k}}{2}}{\sqrt{1-\gamma_{k}}}-1 \right )
+ \frac{1}{4S}(\frac{h}{zJ})^{4}\int\frac{d^{d}k}{(2\pi)^{d}}
\frac{1}{\sqrt{1-\gamma_{k}}}  \nonumber \\
\left\{ \frac{\gamma_{k}}{2}(\tilde{I}_{2}(k)
-\tilde{I}_{3}(k)+\frac{1}{z}\tilde{I}_{1}(k)) -
\frac{(1-\frac{\gamma_{k}}{2})^{2}(1+\tilde{I}_{2}(k))}{1-\gamma_{k}}
+ 1 - \sqrt{1-\gamma_{k}} \tilde{I}_{2}(k) \right\} \;\;,
\end{eqnarray}
where $\tilde{I}_{1}(k)$, $\tilde{I}_{2}(k)$ and $\tilde{I}_{3}(k)$
are given in the appendix. The fact that the second term in the right
hand side of (\ref{qfd}) is positive suggests that disorder
enhances quantum fluctuations.

One can calculate the helicity modulus in a similar way as one does
for the pure model. In the disordered case, the effective Hamiltonian
for phonons become considerably more complicated for a system with
anti-periodic boundary condition. For classical spins, the equations
for the ground state are given by
\begin{eqnarray}\label{ceomap}
sin\theta_{j}\sum_{<j'>}cos\phi_{j'}cos\theta_{j'} = \frac{h_{j}}{J}
cos\phi_{j}cos\theta_{j} \;\; , \nonumber \\
sin\phi_{j}\sum_{<j'>}cos\phi_{j'}cos\theta_{j'} = cos\phi_{j}
\sum_{<j'>}sin\phi_{j'}cos\theta_{j'} \;\; ,
\end{eqnarray}
where the classical spin is defined by $\vec{S}^{classical}_{j} =
S(sin\phi_{j}cos\theta_{j}, sin\theta_{j}, cos\phi_{j}cos\theta_{j})$.
Obviously, we expect the phase twist to be large where the tilt of
the spins away from the ordering plane is large and vice versa. For
a general random field distribution, (\ref{ceomap}) may be solved
numerically. The helicity modulus
\begin{equation}
\gamma(T=0) = J(1 - a(\frac{h}{zJ})^{2} + {\cal O}(h^{4})) \;\;,
\end{equation}
with $a>0$. In the special case where $\{h_{j}\}$ is given by a
bimodal distribution, $P(h_{j}) = \frac{1}{2}(\delta(h_{j}-h_{0}) +
\delta(h_{j}+h_{0}))$, the solution for $\{\phi_{j}\}$ in
(\ref{ceomap}) is the same as that for a uniform field $h_{0}$, and
$a=1$. Since there is no shift in the phonon spectrum to the order
$h^{2}$, there is no quantum correction to $\gamma$ to that order
in the disorder. One might ask if the (low) temperature
dependence of $\gamma$ is affected by the disorder. However, since
the low energy excitation remains to be phonon like, the density
of state $N(\omega) \propto \omega^{d-1}$, assuming that the single
mode approximation remains intact. This implies (see (\ref{gmT})) that
the $T^{d+1}$ behavior of the temperature dependence of the helicity
modulus should be unaltered by the presence of weak disorder.

To summarize our results, we find that to the lowest non-vanishing
order in $h/zJ$, and relative to the pure system, the speed of sound
is reduced ($\delta c_{s} \propto (\frac{h}{zJ})^{4}$), the condensate
density is reduced classically, but unaltered quantum mechanically,
while the superfluid density is reduced classically, but remains
unchanged quantum mechanically. To higher order of $h$, disorder
tends to enhance quantum fluctuations.

So far we considered only a special case where $\overline{h_{j}}
\equiv h_{0} =0$. In general, $h_{0}$ is nonzero and instead of
(\ref{theta}), we have
\begin{equation}
\theta_{j} = \frac{1}{\sqrt{N}} \sum_{k}e^{-ikx_{j}}
\frac{\sqrt{1-H_{0}^{2}}}{1-\gamma_{k} H_{0}^{2}} \frac{h_{k}}{zJ}
+ {\cal O}(h^{2}) \;\;,
\end{equation}
with
$$
h_{k} = \frac{1}{\sqrt{N}}\sum_{j}e^{-ikx_{j}}h_{j} \;\;,
H_{0} = \frac{h_{0}}{zJ} \;\;.
$$
Compare with (\ref{theta}), we see that the primary role of a
non-vanishing $h_{0}$ is to introduce correlations among
$\theta_{j}$'s. Calculation with finite $h_{0}$ is more complicated.
To see physics, however, one can consider the case of correlated
randomness. Thus, we consider (\ref{rfdist}) with
\begin{equation}
\overline{h_{j}} = 0 \;\;,\;\;\; \overline{h_{i}h_{j}} =
f(|x_{i} - x_{j}|) \;\; .
\end{equation}
where $f(x)$ is some arbitrary function. To the lowest order, poles
of the Green function (\ref{Gf}) are given by
\begin{equation}\label{omcf}
\omega^{2} = (\frac{zJ}{S})^{2}(1-\gamma_{k})
(1-\gamma_{k}(\frac{\tilde{h}}{zJ})^{2}) \;\;,
\end{equation}
where $\tilde{h}^{2}$ is the average of a pair of random field at
nearest neighbor, $\overline{h_{j}h_{j+\delta}} = \tilde{h}^{2}$.
Thus, in the weak randomness limit, correlated random fields reduce
the excitation energy, and hence the speed of sound in the same
manner as a uniform field (see (\ref{omegah0})), and do not
induce finite scattering rate for the Bloch phonons at the
lowest order. (Note that in (\ref{omcf}) only the nearest
neighbor correlation in $h_{j}$ matters.)

\section{Discussions}

As remarked previously, our approximation consists of a double
expansion in $1/S$ and the strength of disorder $h$ about
a saddle point solution of (\ref{hr0}) which becomes exact in
the classical ($S=\infty$) limit.
Now we re-examine this approximation scheme in terms of the
original boson Hamiltonian (\ref{hhc}).
In terms of the {\em hard-core} boson operators, the classical
spin ground state described by (\ref{ceom}) corresponds to
a Gutzwiller-type trial wavefunction \cite{MA1}:
\begin{equation}
|\Psi_{G}> = \prod_{j} (sin\varphi_{j} +
cos\varphi_{j}b^{+}_{j})|0> \;\; ,
\end{equation}
with $ \varphi_{j} = \pi /4 - \theta_{j}/2$.
Minimizing the energy $E = <\Psi_{G}|{\cal H}|\Psi_{G}>$ with
respect to $\{ \theta_{j} \}$, one recovers (\ref{ceom}).
In this state, the order parameter
\begin{equation}
b \equiv \frac{1}{N} \overline{<\Psi_{G}|\sum_{j} b_{j} |\Psi_{G}>}
= \frac{1}{2N}\overline{\sum_{j}cos\theta_{j}} \;\; ,
\end{equation}
which is equivalent to the in-plane magnetization of
the classical spins (with $S=1/2$) considered previously.
The quantum fluctuations described by the spin-wave theory
correspond to the Gaussian fluctuations about this Gutzwiller
state.  Within this scheme, our perturbative calculation
shows that the low energy Bloch modes are rather robust,
and disorder has little effect on the long wavelength quantum
mechanical behavior of the system. We cannot, however, rule out
the possibility of a stronger effect of disorder (e.g. a $h^{2}$
correction to the speed of sound) as one goes to higher orders of
$1/S$. At higher orders of $1/S$, phonon-phonon scatterings
take place and cause decay of these quasi-particles.
Presumably such decay processes are stronger than those of
the Bloch phonons in the corresponding pure system due to
disorder enhanced quantum fluctuations. How
does disorder enhance these fluctuation effects is an
important question that yet to be investigated.

Our perturbative study shows that superfluidity is rather
insensitive to the presence of weak disorder.
This result is fully in agreement with what one would expect
intuitively: since superfluidity is due to quantum coherence
on the macroscopic scale, superfluid should be rigid against
weak impurity scatterings. A similar statement for fermion systems
is expressed in the content of the Anderson Theorem for the BCS
superconductors with weak homogeneous non-magnetic impurities
\cite{AT}. A natural question to be asked is then how would
the system evolves with increasing randomness.
At zero temperature, as disorder becomes stronger, or as
effects of interaction and quantum zero point fluctuations
become more important, or both, one may reach a point where
a transition from the superfluid phase to a disordered (Bose
glass) phase takes place. Since in the Bose glass phase the
low energy excitations are single particle like, one may
expect that the speed of sound, which characterizes the low
energy excitations of the superfluid phase, vanishes at that
point. Thus the transition point can be located naively by
setting $c_{s}$ in (\ref{csr}) to zero, which gives
\begin{equation}
\frac{1}{2}(\frac{h}{zJ})^{4}A(0) =1 \;\;.
\end{equation}
Of course, long before this point is reached our (weak
disorder)  approximation loses its legitimacy. However, it
is still a non-trivial and interesting question that whether
the Gaussian fluctuations contained in the first order $1/S$
expansion are capable to describe such a phase transition.
Since this approach is based upon an expansion about the
ordered phase, phase transition is expected to be signaled
by instabilities of the expansion, such as appearance of
negative energy modes or strong divergence of quantum zero
point fluctuation corrections.
The present weak disorder calculation cannot
answer such questions concerning strong disorder, but this
approach does provide a scheme for further investigations.
Since the effective Hamiltonian is quadratic to the first
order of $1/S$, exact numerical diagonalizations are possible
for finite systems upto sizes which are unreachable otherwise.
Thus it provides us a way of study the low energy
excitation spectrum in the strong disordered system.
This will be discussed in a forth coming work \cite{PMZ}.
Of course, such a study can only address these questions
within the Gaussian scheme. Other than simply having a
vanishing speed of sound, there are different possibilities
for the phonon mode to evolve into the single
particle continuum as the system is tuned into the
disordered phase. For instance, $c_{s}$ may remain finite
while phonon decays strongly with increasing disorder so
that its spectral weight vanishes at the transition point.
How precisely the phonon mode evolves with increasing
randomness is still an open question which can only be
answered by going beyond the Gaussian approximation.

In conclusion, we have found that the spin representation of
the boson problem is an effective approach for investigating
the effect of disorder. Within the Gaussian approximation,
our perturbative study shows that the superfluidity remains
robust in the presence of weak disorder, while random fields
have scatterings with Bloch phonons with a rate
proportional to $\omega ^{d+1}$ and give a weak reduction
of the speed of sound. Our calculations also suggest that
stronger disorder tends to enhance quantum fluctuations
which may eventually cause the destruction of
superfluidity \cite{MHL}.

\begin{center}
{\bf Acknowledgement}
\end{center}

The author would like to thank M. Ma for numerous valuable
discussions and comments, and for his critical reading of the
manuscript. This work was supported in part
by the National Science Foundation under Grant No. DMR-9101542.

\appendix{}

In this appendix we evaluate the self energy $\Sigma^{*}(k,\omega)$
to order $h^{4}$. Assuming $h$ is small, one can solve (18)
perturbatively. To the lowest order of $h_{j}$, $\theta_{j}$ is
simply given by $H_{j} \equiv h_{j}/zJ$. To the next order, one has
\begin{equation}\label{theta}
\theta_{j} = H_{j} (1 - \frac{1}{3}H^{2}_{j} +
\frac{1}{2z}\sum_{<j'>}H^{2}_{j'} + \ldots )\;\; .
\end{equation}
Here we see that $\theta_{j}$ starts to couple to the
field at the neighbor sites as one goes to higher order
of $h_{j}$. We calculate the
quantities
\begin{equation}
\overline{u_{0}} = \frac{zJ}{4S}(\overline{H^{4}} -
(\overline{H^{2}})^{2}) \;\; ,
\end{equation}
\begin{equation}
\overline{u_{k}u_{-k}} = (\frac{zJ}{S})^{2} \frac{1}{4N}
(1-\gamma_{k})^{2}(\overline{H^{4}} - (\overline{H^{2}})^{2}) \;\; ,
\end{equation}
\begin{equation}
\overline{(v_{k,-k'} + v_{-k',k})(v_{k',-k} + v_{-k,k'})} =
(\frac{zJ}{S})^{2}
\frac{1}{4zN} (1 + \gamma_{k+k'}) (\overline{H^{2}})^{2} \;\; ,
\end{equation}
where $\overline{H^{n}}$ is the $n$-th moment of $H_{j}$.
With these results, the self energy
\begin{equation}
\Sigma^{*}(k,\omega) \equiv
\left ( \begin{array}{cc}
\Sigma_{11}^{*}(k,\omega) & \Sigma_{12}^{*}(k,\omega)  \\
\Sigma_{21}^{*}(k,\omega) & \Sigma_{22}^{*}(k,\omega)
\end{array} \right ) \;\; ,
\end{equation}
is evaluated as
\begin{eqnarray}
\Sigma_{11}^{*}(k,\omega) = \frac{zJ}{4S}
(\overline{H^{4}}-(\overline{H^{2}})^{2}) +
(\frac{zJ}{2S})^{2}\frac{1}{N}(\overline{H^{4}} -
(\overline{H^{2}})^{2} )
\sum_{k'} [ (\omega + \frac{zJ}{S}(1 - \frac{\gamma_{k'}}{2}))
\nonumber \\
\frac{(1 - \gamma_{k-k'})^{2}}{D^{0}(k',\omega)}  ]
+ (\frac{zJ}{2S})^{2}\frac{2}{zN}(\overline{H^{2}})^{2}
\sum_{k'}\frac{zJ}{S}(1-\gamma_{k'})\frac{1 +
\gamma_{k-k'}}{D^{0}(k',\omega)}   \;\; ,
\end{eqnarray}
\begin{eqnarray}
\Sigma_{22}^{*}(k,\omega) = -\frac{zJ}{4S} (\overline{H^{4}} -
(\overline{H^{2}})^{2}) +
(\frac{zJ}{2S})^{2}\frac{1}{N}(\overline{H^{4}} -
(\overline{H^{2}})^{2})
\sum_{k'} [ (\omega - \frac{zJ}{S}(1 - \frac{\gamma_{k'}}{2}))
\nonumber \\
\frac{(1 - \gamma_{k-k'})^{2}}{D^{0}(k',\omega)} ]
- (\frac{zJ}{2S})^{2}\frac{2}{zN}(\overline{H^{2}})^{2}
\sum_{k'}\frac{zJ}{S}(1-\gamma_{k'})\frac{1 + \gamma_{k-k'}}
{D^{0}(k',\omega)}  \;\; ,
\end{eqnarray}
\begin{eqnarray}
\Sigma_{12}^{*}(k,\omega) = -\Sigma_{21}^{*}(k,\omega) =
(\frac{zJ}{2S})^{2}\frac{1}{N}(\overline{H^{4}} -
(\overline{H^{2}})^{2})
\sum_{k'} \frac{zJ}{S}\frac{\gamma_{k'}}{2}
\frac{(1 - \gamma_{k-k'})^{2}}{D^{0}(k',\omega)} \nonumber \\
- (\frac{zJ}{2S})^{2}\frac{2}{zN}(\overline{H^{2}})^{2}
\sum_{k'}\frac{zJ}{S}(1-\gamma_{k'})\frac{1 + \gamma_{k-k'}}
{D^{0}(k',\omega)}  \;\; ,
\end{eqnarray}
with $D^{0}(k,\omega)$ given by (\ref{det0}). Since
\begin{equation}
(\tilde{F}^{0-1} - \Sigma^{*})^{-1} = \frac{1}{D(k,\omega)}
\left ( \begin{array}{cc}
\omega + \frac{zJ}{S}(1 - \frac{\gamma_{k}}{2}) - \Sigma^{*}_{22} &
- (\frac{zJ}{S}\frac{\gamma_{k}}{2} - \Sigma^{*}_{12}) \\
(\frac{zJ}{S}\frac{\gamma_{k}}{2} + \Sigma^{*}_{21}) &
\omega - \frac{zJ}{S}(1 - \frac{\gamma_{k}}{2}) - \Sigma^{*}_{11}
\end{array} \right ) \;\; ,
\end{equation}
the poles in $G(k,\omega)$ is given by the zero's in
\begin{eqnarray}\label{deta}
D(k,\omega)\equiv det(\tilde{F}^{0-1}-\Sigma^{*})=\omega^{2}\left[
1 - \frac{1}{2} (\frac{zJ}{S})^{2}(\overline{H^{4}} -
(\overline{H^{2}})^{2})I_{2}(k,\omega) \right ] \nonumber \\
- (\frac{zJ}{S})^{2}(1 - \gamma_{k}) -\frac{1}{2}(\frac{zJ}{S})^{2}
(\overline{H^{4}}-(\overline{H^{2}})^{2})(1 - \frac{\gamma_{k}}{2})
(1 + (\frac{zJ}{S})^{2}I_{2}(k,\omega)) \nonumber \\
+\frac{1}{2}(\frac{zJ}{S})^{4}(\overline{H^{4}}-
(\overline{H^{2}})^{2})(1 - \gamma_{k})I_{3}(k,\omega)
-\frac{1}{z}(\frac{zJ}{S})^{4}(\overline{H^{2}})^{2}
(1 - \gamma_{k})I_{1}(k,\omega) \;\; ,
\end{eqnarray}
with
\begin{equation}
I_{1}(k,\omega) = \frac{1}{N}\sum_{k'}
\frac{(1-\gamma_{k'})(1+\gamma_{k-k'})}{D^{0}(k',\omega)} \;\;  ,
\end{equation}
\begin{equation}
I_{2}(k,\omega) = \frac{1}{N}\sum_{k'}
\frac{(1 - \gamma_{k-k'})^{2}}{D^{0}(k',\omega)} \;\; ,
\end{equation}
\begin{equation}
I_{3}(k,\omega) = \frac{1}{N}\sum_{k'}\frac{\gamma_{k'}}{2}
\frac{(1 - \gamma_{k-k'})^{2}}{D^{0}(k',\omega)} \;\; .
\end{equation}
When the distribution of random fields is Gaussian,
\begin{equation}
\overline{H^{4}} - (\overline{H^{2}})^{2} = 2(\frac{h}{zJ})^{4} \;\; .
\end{equation}
To the order ${\cal O}(h^{4})$, one can substitute $\omega$
in $I_{j}$ ($j = 1,2,3$) by $\omega_{k}^{0}$. Since we are
only interested in the case where $\omega > 0$, we may ignore
the singularities for negative $\omega$. We have
\begin{equation}\label{poleG}
\omega^{2} = \omega^{02}_{k} \left [1 + (\frac{h}{zJ})^{4} \left (
\tilde{I_{2}}(k) + (1 - \frac{\gamma_{k}}{2})
\frac{1+\tilde{I_{2}}(k)}{1-\gamma_{k}} - \tilde{I_{3}}(k) +
\frac{1}{z}\tilde{I_{1}}(k) \right ) \right ] \;\; ,
\end{equation}
where ($\eta \rightarrow 0^{+}$)
\begin{equation}
\tilde{I_{1}}(k) = \frac{1}{N}\sum_{k'} \frac{(1-\gamma_{k'})
(1+\gamma_{k-k'})}{\gamma_{k'} - \gamma_{k} + i\eta} \;\; ,
\end{equation}
\begin{equation}
\tilde{I_{2}}(k) = \frac{1}{N}\sum_{k'}
\frac{(1 - \gamma_{k-k'})^{2}}{\gamma_{k'}-\gamma_{k}+i\eta} \;\;  ,
\end{equation}
\begin{equation}
\tilde{I_{3}}(k) = \frac{1}{N}\sum_{k'}\frac{\gamma_{k'}}{2}
\frac{(1 - \gamma_{k-k'})^{2}}{\gamma_{k'} - \gamma_{k} + i\eta} \;\;  .
\end{equation}
To the order of $\omega^{2}$ and $k^{2}$,
\begin{equation}\label{polelw}
\omega^{2} = \omega^{02}_{k} \left (1 - (\frac{h}{zJ})^{4}
(\frac{3z+3}{2z} + \frac{1}{N}\sum_{k'}\frac{\frac{1}{d^{2}}\sum_{j=1}
^{d}sin^{2}k'_{j}}{1-\gamma_{k'} - i\eta}) \right ) \;\; ,
\end{equation}
which implies a correction to the speed of sound due to disorder
shown in eqn. (\ref{csr}).
The imaginary part of (\ref{deta}) is of the order $\omega^{d+2}$
, which can be seen through a simple dimensional analysis. This
gives the life time of phonons $\propto \omega^{-(d+1)}$.

To study the effect of disorder on the quantum corrections to the
order parameter (\ref{qfd}), one needs to compute
\begin{equation}
\overline{<a^{+}_{j}a_{j}>} = \frac{1}{N} \sum_{k}
\overline{<a^{+}_{k}a_{k}>} = \frac{i}{N} \sum_{k}\int
\frac{d\omega}{2\pi}E_{1}^{+}G(k,\omega)e^{i\omega 0^{+}} \;\; .
\end{equation}
Using the result for $G$ obtained above, and completing the
$\omega$ integral by the usual contour integration over the
semicircle at the lower half complex plane, we obtain
\begin{eqnarray}\label{dm}
\overline{<a^{+}_{j}a_{j}>} = \frac{1}{2}\int
\frac{d^{d}k}{(2\pi)^{d}}
\left (\frac{1-\frac{\gamma_{k}}{2}}{\sqrt{1-\gamma_{k}}}-1 \right )
+ \frac{1}{4}(\frac{h}{zJ})^{4}\int\frac{d^{d}k}{(2\pi)^{d}}
\frac{1}{\sqrt{1-\gamma_{k}}}  \nonumber \\
\left \{ \frac{\gamma_{k}}{2}(\tilde{I}_{2}(k)
-\tilde{I}_{3}(k)+\frac{1}{z}\tilde{I}_{1}(k)) -
\frac{(1-\frac{\gamma_{k}}{2})^{2}(1+\tilde{I}_{2}(k))}{1-\gamma_{k}}
+ 1 - \sqrt{1-\gamma_{k}} \tilde{I}_{2}(k) \right \} \;\;.
\end{eqnarray}
One can verify numerically that the second term is positive.
{}From this result, it is easy to show that the relative
quantum correction to the order parameter is given by (\ref{qfd}).


\begin{references}
\bibitem{MA1}
This interpretation of the classical spin ground state was
pointed out to the author by M. Ma.
\end{references}
\end{document}